# MECHANISM OF THERMAL CONDUCTIVITY REDUCTION IN FEW-LAYER GRAPHENE

Dhruv Singh    Jayathi Y. Murthy    Timothy S. Fisher

School of Mechanical Engineering and Birck Nanotechnology Center

Purdue University, West Lafayette, IN-47907, USA

## ABSTRACT

Using the linearized Boltzmann transport equation and perturbation theory, we analyze the reduction in the intrinsic thermal conductivity of few-layer graphene sheets accounting for all possible three-phonon scattering events. Even with weak coupling between layers, a significant reduction in the thermal conductivity of the out-of-plane acoustic modes is apparent. The main effect of this weak coupling is to open many new three-phonon scattering channels that are otherwise absent in graphene. However, reflection symmetry is only weakly broken with the addition of multiple layers, and ZA phonons still dominate thermal conductivity. We also find that reduction in thermal conductivity is mainly caused by lower contributions of the higher-order overtones of the fundamental out-of-plane acoustic mode. The results compare remarkably well over the entire temperature range with measurements of graphene and graphite.

## I. INTRODUCTION

Since the discovery of graphene and its remarkable electrical [1,2] and thermal properties [3-5], scalability issues with mechanical exfoliation have led to many studies of its properties when in contact with a substrate [6-8]. Bilayer and few-layer graphene have been investigated and shown to exhibit a tunable band gap [9-11]. Graphene on a substrate has shown significantly lower thermal conductivity compared to its suspended counterpart [12] that is believed to be caused by the suppression of thermal transport in the out-of-plane acoustic modes. At the same time, measurements of suspended single-layer graphene and carbon nanotubes have consistently shown values of thermal conductivity higher than graphite [3, 4, 13]. On the other hand, the use of carbon nanotubes as thermal interface materials and in suspensions has posed great challenges due to high contact resistance between individual nanotubes [14, 15] – a result of the weak coupling between nanotubes [16, 17]. It is a result of this weak coupling that graphite has extremely low thermal conductivity, elastic constants and sound velocity perpendicular to the layers [18]. Two recent sets of experimental measurements on thermal conductivity of few-layer graphene are of particular interest here [19, 20]. Reference [19] considers the dimensional transition of thermal conductivity from single layer graphene to graphite in suspended samples by systematic measurement with respect to the number of layers. The measurements reveal that in-plane thermal conductivity decreases as the number of layers increases and saturates to a constant value beyond four layers. Data from reference [20] focuses on multilayered graphene encased between $SiO_2$ substrates. Due to interactions with the substrate, a strong reduction is seen in thermal conductivity as compared to suspended graphene, and the measurements indicate that thermal conductivity increases as layers are added – a trend opposite to that observed in suspended samples. These experiments reveal that the effective thermal conductivity increases with the number of layers as the strength of interaction with the $SiO_2$ substrate decreases with depth into the graphene film. *In this paper, we theoretically analyze the reduction in intrinsic thermal conductivity of suspended few-layer graphene samples to understand the transition in thermal conduction from single graphene to graphite.*

A wealth of experimental data [3-5, 12, 19-22] and rigorous theoretical calculations [12, 23-25] suggest that thermal conductivity in graphene is dominated by the out-of-plane acoustic (ZA) phonons with a relatively small contribution from the in-plane acoustic (LA/TA) modes. It is now understood that use of Klemens-like relaxation time expressions [26, 27] to describe phonon scattering processes in graphene and carbon nanotubes are inadequate in describing thermal transport [28]. More detailed models that account for the admissible phonon interactions in graphene but retain the Klemens approximations for matrix elements, such as those reported in [19, 29], also suffer from inaccurate descriptions of thermal conduction by ZA phonons for the following reasons:

- The selection rule that arises out of the reflection symmetry of the graphene layer is not present in these expressions. In fact, the long wavelength approximation (LWA) of the matrix elements for specific interactions is itself responsible for many errors especially when the scattering involves the ZA modes.
- Such expressions when applied to few-layer graphene do not accurately account for phonon degeneracy. Phonon dispersion curves of single and *N* layer graphene are degenerate throughout most of the Brillouin zone except near the zone center. This means that allowed three-phonon scattering processes (satisfying energy and momentum conservation) in *N* layer graphene increase by factor of $\sim N^2$ in terms of scattering rate for each phonon branch, which implies a drastic reduction in the *intrinsic* thermal conductivity with the addition of every new layer. The flaw of this argument resides in the assumption that the strength of scattering processes involving vibrational modes of different layers is the same.

While the precise nature of interlayer bonding in graphite remains an active research subject [30], and is notoriously difficult to capture through first-principles simulations [31], the aforementioned assumption is questionable since the interatomic forces between different layers in graphite are very weak compared to the in-plane interactions and one expects thermal conductivity behavior to be similar to single layer graphene.

Experimentally, the highest reported difference between the thermal conductivity of single-layer graphene and high quality bulk graphite remains within a factor of 2 [3, 4]. This clearly suggests that the appearance of many more admissible interactions in graphite (due to the existence of the $\Gamma$-A $k$ space), does not decrease the thermal conductivity in proportion to the number of layers. In this paper, we use a direct approach based on empirical interatomic potentials to compute thermal conductivity in single- and few-layer *AA* stacked graphene. We show that any noticeable changes in the phonon dispersion curves of single and few-layer graphene are limited to regions near the $\Gamma$ point. We also show that the effect of interlayer coupling on anharmonicity is to open new phonon scattering channels involving an odd number of out-of-plane phonons, with the ZA→ZA+ZA phonon scattering channel being the most resistive. These processes do not contribute any thermal resistance in single layer graphene, but are responsible for the decrease in the intrinsic thermal conductivity of the ZA branch in few-layer graphene; the thermal conductivity contributions of other modes are found to be relatively unaffected. Using a solution of the linearized phonon Boltzmann transport equation (without resorting to Klemens' matrix elements and the single-mode relaxation time approximation) we clearly show how the transition in thermal conductivity occurs from single layer to graphite.

**II. PHONON DISPERSION IN FEW-LAYER GRAPHENE**

Phonon frequencies and polarization vectors can be computed from the eigenvalue problem,

$$\omega^2(\vec{k})e_{\alpha,m}(\vec{k}) = \sum_{n,\beta}(m_m m_n)^{-1/2}\sum_i \phi_{\alpha\beta}^{m(0)n(i)} \exp(-i(\vec{k}\cdot\vec{R}_i - \omega t))e_{\beta,n}(\vec{k}) \qquad (1)$$

where, $\phi_{\alpha\beta}^{m(0)n(i)}$ is the harmonic interatomic force constant (IFC) between atoms $m$ (in the *reference* unit cell) and $n$ (in the $i^{th}$ unit cell). $e_{\alpha,m}(\vec{k})$ is the $\alpha^{th}$ component of the polarization vector corresponding to the basis atom $m$. Indices $m$ and $n$ run from 1 to $2N$ in $N$ layer graphene. $\vec{R}_i$ is the translational vector connecting the $i^{th}$ unit cell to the reference unit cell. The force fields are described using the Tersoff interatomic potential for in-plane interactions (with the parameterization in [32]). The Lennard-Jones (LJ) potential is used to model forces between atoms belonging to different layers. The parameters for the LJ potential used here are $\varepsilon = 0.0024$ eV and $\sigma = 3.41$ Å which successfully reproduces the interlayer cohesion energy and the *c*-axis compressibility of graphite [33]. The harmonic and anharmonic IFCs are calculated using central differences on the total crystal energy by systematic displacement of atoms. The procedure employed ensures that these satisfy translational invariance [34]. All the derivatives are calculated at the equilibrium lattice constants $a$ (in-plane) and $c_0$ (interlayer distance) which are arrived at by energy minimization for each structure.

Computed phonon dispersion curves along the $\Gamma$-$M$ direction are shown in Figure 1 (b), (c) and (d) for 1, 2 and 4-layer graphene respectively. The unit cell for $N$-layer graphene consists of $N$ multiples of a 2-atom basis ( as for graphene). The Brillouin zone (BZ) geometry remains the same as that for graphene, but there are $6N$ phonon branches in $N$ layer graphene. The salient differences may be understood by examining the dispersion curves of bilayer graphene. Throughout most of the BZ, the phonon branches are degenerate. A splitting of the ZA phonon branch is apparent near the $\Gamma$ point (labeled as $ZA_2$). At the $\Gamma$ point, the highest frequency of the out-of-phase $ZA_2$ mode is 77.2, 98.1 and 105.9 cm$^{-1}$ for 2, 3 and 4 layers respectively. It is remarkable that without any fitting of the LJ parameters, the obtained $\Gamma$ point frequency of the $ZA_2$ mode is in excellent agreement with recently published first-principles calculations of phonons in few-layer graphene [35]. Since the interlayer coupling is very weak, the in-plane interatomic force constants are relatively unaffected, which indicates that the splitting of LA/TA phonon modes at the $\Gamma$ point is much lesser in extent than that of ZA phonons. Figure 1(a) also shows the atomic displacements corresponding to the highest overtone of the fundamental ZA phonon mode at the $\Gamma$ point for bilayer and 4 layer graphene. This mode corresponds to out-of-phase vibrations of adjacent layers (labeled $ZA_2$ for bilayer and $ZA_4$ for 4 layer graphene).

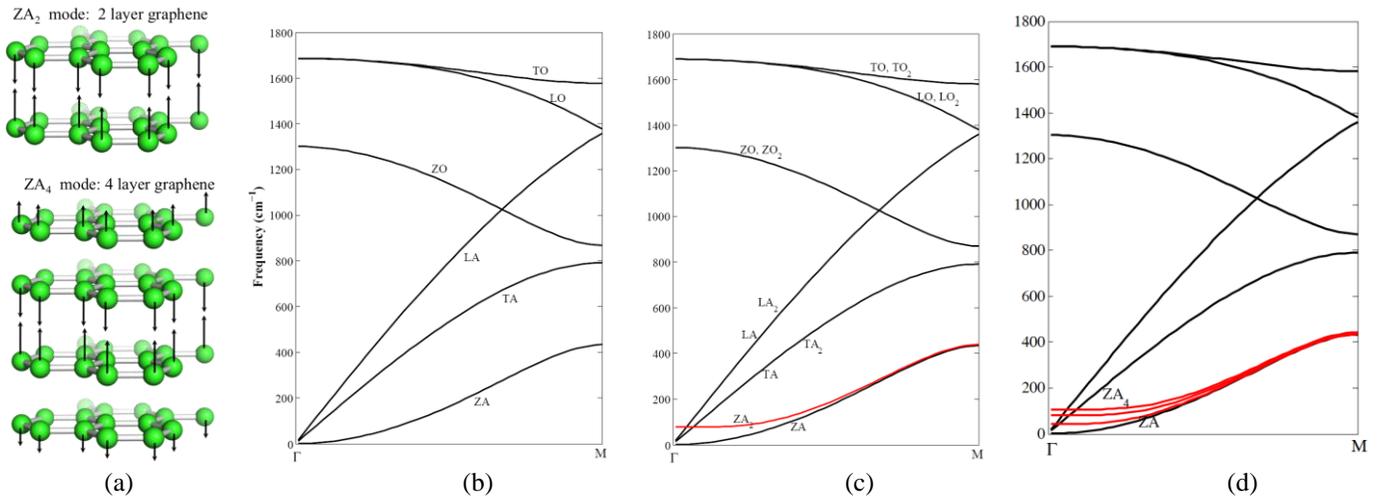

(a)  (b)  (c)  (d)

Figure 1 *(a)Phonon eigenvector for the highest overtone of the ZA mode in 2 and 4 layer graphene (b) Phonon dispersion for single layer graphene (c) Phonon dispersion curves for 2 layer graphene (d) Phonon dispersion curves for 4 layer graphene. The splitting of dispersion curves due to interlayer interaction is significant in these figures only for ZA modes but exists for all modes.*

At an arbitrary wave vector different from the $\Gamma$ point, there is a slight mixing between the in-plane and out-of-plane vibrational modes. The C-C bond-length in few-layer graphene changes only slightly (1.4388 Å in single layer graphene to 1.4382 Å in 4-layer graphene) with the introduction of LJ coupling between layers. Here $c_0$ is the interlayer distance between two graphene sheets (3.43 Å in bilayer graphene to 3.41 Å in 4-layer graphene). To present a consistent set of results, all thermal properties are reported after division by $Nc_0 = N*3.41$ Å (this is done in order to facilitate an easy comparison to bulk graphite and to maintain consistency).

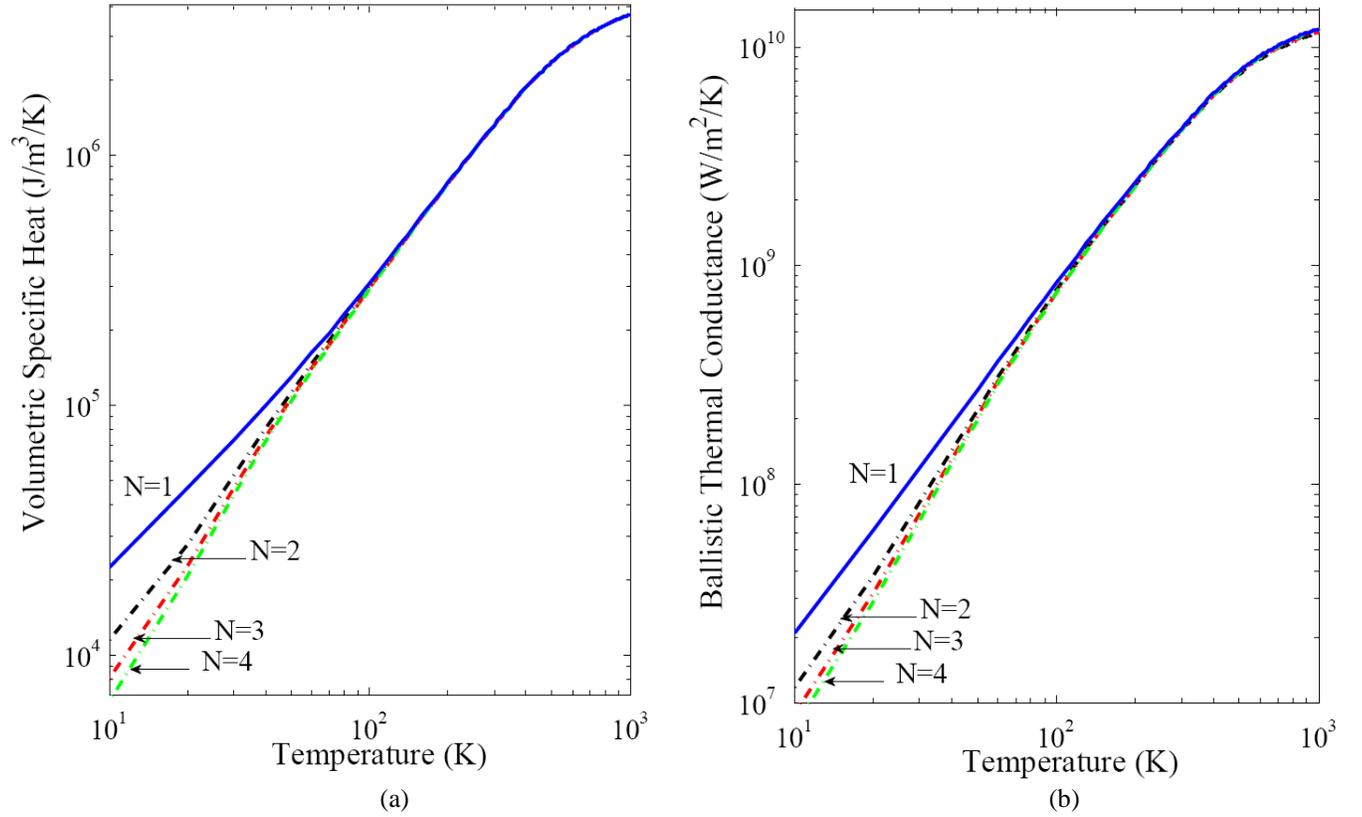

(a)           (b)

Figure 2 (a) *Volumetric specific heat capacity of 1-4 layer graphene (b) Ballistic thermal conductance of 1-4 layer graphene along the $\Gamma$-M direction*

The volumetric specific heat of few-layer graphene can be calculated as,

$$C_v = \frac{1}{V}\frac{\partial}{\partial T}\sum_{p,\vec{k}} \hbar\omega_p n^0(\omega_p,T) = \frac{1}{Nc_0}\sum_{p=1..6N}\frac{\partial}{\partial T}\int \frac{\hbar\omega_p}{e^{\hbar\omega_p/k_BT}-1}\frac{dk_x dk_y}{(2\pi)^2} \qquad (2)$$

where $N$ is the number of layers, $n^0(\omega_p, T)$ is the Bose-Einstein distribution at temperature $T$. Figure 2(a) shows the variation of specific heat with temperature for 1-4 layer *AA* stacked graphene sheets. At low temperatures(<50K), a largedifference can be seen between the specific heat of single and few-layer graphene sheets but this difference quickly decreases to less than 1% at room temperature. From the knowledge of phonon dispersion (Figure 1), it can be understood that this difference arises primarily from the splitting of the fundamental acoustic modes and the fact that at very low temperatures these higher order overtones are not thermally active. However, since the frequencies of these overtones is low enough (< 100 cm$^{-1}$), they start to show significant occupation at temperatures 100 K and higher. Phonon dispersion curves can also be used to calculate the ballistic thermal conductance of these sheets. The conductance $G$ along any direction $\vec{n}$ may be calculated using,

$$G = \frac{1}{V}\frac{\partial}{\partial T}\sum_{\substack{p,\vec{k}\\ \vec{v}\cdot\vec{n}\geq 0}} \hbar\omega_p(\vec{v}\cdot\vec{n})n^0(\omega_p,T) = \frac{1}{Nc_0}\sum_{p=1..6N}\frac{\partial}{\partial T}\int_{\vec{v}\cdot\vec{n}\geq 0} \frac{\hbar\omega_p \vec{v}\cdot\vec{n}}{e^{\hbar\omega_p/k_B T}-1}\frac{dk_x dk_y}{(2\pi)^2} \qquad (3)$$

The ballistic thermal conductance along the direction *Γ-M* as a function of temperature is plotted in Figure 2(b). Clearly, any differences in thermal conductance between single and few-layer graphene is restricted to below 100K. At temperatures above 100 K, the branch-wise contribution to thermal conductivity and ballistic thermal conductance remains very similar to single layer graphene. Therefore, we conclude that any differences in thermal conductivity at room temperature and higher should not be attributed to changes in phonon group velocity or mode specific heat, as variations these quantities with respect to single-layer graphene are limited to low temperatures. However, phonon scattering rates can differ significantly going from single layer to few-layer graphene, and differences in thermal conductivity may result.

### III. PHONON SCATTERING, LINEARIZED BTE AND THERMAL CONDUCTIVITY

Thermal conductivity of few-layer graphene sheets can be calculated by solving the phonon Boltzmann transport equation under weak nonequilibrium conditions (in this case a small temperature gradient). At steady state, the linearized phonon Boltzmann transport equation (under the presence of a small temperature gradient) can be rewritten to form an equation set for the deviation $\vec{\Psi}_{\vec{k}(p)}$ from equilibrium of the phonon population $n^0_{\vec{k}(p)}$ for a phonon of polarization $p$ and wavevector $\vec{k}$ (denoted as $\vec{k}(p)$), dependent upon those of the interacting modes [24, 37-40],

$$\vec{\Psi}_{\vec{k}(p)} = \frac{\hbar \omega_{\vec{k}(p)} n^0_{\vec{k}(p)}\left(n^0_{\vec{k}(p)}+1\right)}{\Gamma_{\vec{k}(p)} T} \vec{v}_{\vec{k}(p)} + \frac{A}{2\pi\hbar^2 \Gamma_{\vec{k}(p)}} X \left\{ \begin{array}{l} \sum_{p',p''} \int n^0_{\vec{k}(p)} n^0_{\vec{k}'(p')} \left(n^0_{\vec{k}''(p'')}+1\right)\left(\vec{\Psi}_{\vec{k}''(p'')} - \vec{\Psi}_{\vec{k}'(p')}\right) \left|\mathfrak{S}_{\vec{k}(p)+\vec{k}'(p') \leftrightarrow \vec{k}''(p'')}\right|^2 \frac{dk'_l}{|\vec{v}_n|} \\ + \frac{1}{2} \sum_{p',p''} \int n^0_{\vec{k}(p)} \left(n^0_{\vec{k}'(p')}+1\right)\left(n^0_{\vec{k}''(p'')}+1\right)\left(\vec{\Psi}_{\vec{k}''(p'')} + \vec{\Psi}_{\vec{k}'(p')}\right)\left|\mathfrak{S}_{\vec{k}(p) \leftrightarrow \vec{k}'(p')+\vec{k}''(p'')}\right|^2 \frac{dk'_l}{|\vec{v}_n|} \end{array} \right\}$$

……(4)

which takes into account the net change in phonon population of this mode through intrinsic type 1 ($\vec{k}(p)+\vec{k}'(p') \leftrightarrow \vec{k}''(p'')$), type 2 ($\vec{k}(p) \leftrightarrow \vec{k}'(p')+\vec{k}''(p'')$) three phonon scattering events and scattering due to sample boundaries (using a boundary scattering rate $\tau^{-1}_{B,\vec{k}(p)}$). The first term on the rhs of Eq. (4), $\vec{\Psi}^0_{\vec{k}(p)}$, depends only on the equilibrium population of the interacting phonon modes (through the quantity $\Gamma_{\vec{k}(p)}$) rather than the nonequilibrium population ($\propto \vec{\Psi}_{\vec{k}(p)}$). $\vec{\Psi}^0_{\vec{k}(p)}$ is the shift in phonon distribution under the single mode relaxation time approximation. The quantity $\Gamma_{\vec{k}(p)}$ has been called the scattering amplitude and is calculated as,

$$\Gamma_{\vec{k}(p)} = \frac{n^0_{\vec{k}(p)}\left(n^0_{\vec{k}(p)}+1\right)}{\tau_{B,\vec{k}(p)}} + \frac{A}{2\pi\hbar^2}\left\{ \sum_{p',p''} \int n^0_{\vec{k}(p)} n^0_{\vec{k}'(p')} \left(n^0_{\vec{k}''(p'')}+1\right) \left|\mathfrak{S}_{\vec{k}(p)+\vec{k}'(p') \leftrightarrow \vec{k}''(p'')}\right|^2 \frac{dk'_l}{|\vec{v}_n|} + \frac{1}{2} \sum_{p',p''} \int n^0_{\vec{k}(p)} \left(n^0_{\vec{k}'(p')}+1\right)\left(n^0_{\vec{k}''(p'')}+1\right)\left|\mathfrak{S}_{\vec{k}(p) \leftrightarrow \vec{k}'(p')+\vec{k}''(p'')}\right|^2 \frac{dk'_l}{|\vec{v}_n|} \right\}$$

………(5)

The integration in Eq. (4) and Eq. (5) is performed along $k_l'$ the line segment corresponding to $\omega+\omega'(p')-\omega''(p'')=0$, which eliminates the delta function with the use of $\vec{v}_n = \nabla_{k''} \omega_{k''p''} - \nabla_{k'} \omega_{k'p'}$ [36]. To preserve accuracy, the search for valid phonon scattering events is performed for every combination of $p, p'$ and $p''$ without using the degeneracy of phonon branches. More details on the methodology used to construct of these line segments is outlined in [24]. $\mathfrak{S}_{\vec{k}(p)+\vec{k}'(p') \leftrightarrow \vec{k}''(p'')}$ contains terms from the anharmonic IFC tensor and using the symmetry with respect to $\vec{k}$, $\vec{k}'$ and $\vec{k}''$, it can be simplified to [23, 37, 38, 40],

$$\mathfrak{S}_{\vec{k}(p)+\vec{k}'(p') \leftrightarrow \vec{k}''(p'')} = -i\left(\frac{\hbar}{2}\right)^{3/2}\sqrt{\frac{1}{\omega_{\vec{k}(p)}\omega_{\vec{k}'(p')}\omega_{\vec{k}''(p'')}}} \sum_l \sum_{m,i} \sum_{n,j} \sum_{\alpha\beta\gamma} \phi^{l(0)m(i)n(j)}_{\alpha\beta\gamma} \frac{e_{\alpha,l}(\vec{k}(p))e_{\beta,m}(\vec{k}'(p'))e_{\gamma,n}(\vec{k}''(p''))}{\sqrt{M_l M_m M_n}} \exp(i\vec{k}'\cdot\vec{R}_i)\exp(i\vec{k}''\cdot\vec{R}_j)$$

……(6)

where $\phi^{l(0)m(i)n(j)}_{\alpha\beta\gamma}$ is the anharmonic third-order interatomic force constant. The summation in Eq. (6) is over all the basis atoms $l, m, n$ (=1…6N), in unit cells 0 (reference), $i$ and $j$ respectively and over the direction indices $\alpha, \beta, \gamma$ (= x, y, z). The set of equations specified by Eq. (4) may be solved iteratively to obtain the thermal conductivity tensor as,

$$\kappa_{\alpha\beta} = \sum_p \int n^0_{\vec{k}(p)}\left(n^0_{\vec{k}(p)}+1\right)\frac{\hbar\omega}{k_B T} v_{\vec{k}(p),\alpha} \Psi_{\vec{k}(p),\beta} \frac{dk_x dk_y}{(2\pi)^2}; \quad \alpha,\beta = x,y \qquad (7)$$

It should be pointed out that using this approach, one cannot directly calculate the out-of-plane thermal conductivity as crystal periodicity is considered only along *x* and *y* (i.e., in-plane) directions. The details of our computational procedure may be found in [24].

**IV. RESULTS AND DISCUSSION**

The computed thermal conductivity using the single mode relaxation time approximation (SMRT) for 1-4 layer graphene sheets is shown in Figure 3(a). A significant decrease is seen as we move from single- to 2-layer graphene. The extent of decrease in thermal conductivity lessens as more layers are added and eventually saturates by 4-layer graphene, with more pronounced effects at low temperatures. As noted earlier, the detailed scattering interactions and the low intrinsic scattering levels in carbon nanotubes [25, 28] and graphene [23, 24] render the SMRT grossly inadequate in describing thermal conductivity in these materials. Nevertheless, the SMRT gives a good estimate of the net thermal resistance offered by the extra scattering channels that open in few-layer graphene (while it does not correctly account for N processes). The drop in thermal conductivity (under the SMRT) with respect to single-layer graphene is approximately 27% for few layer graphene with N=2-4 (at 300 K).

Figure 3(b) shows the thermal conductivity of 1-4 layer graphene sheets using an iterative solution of the BTE, which rigorously accounts for all the N and U scattering processes and their dependence on the non-equilibrium phonon populations of the phonon modes. At room temperature, the values of thermal conductivity for single layer graphene are approximately 3.3 times higher than that computed using SMRT. The percentage difference of the total thermal conductivity compared to single layer suspended graphene at 300 K is 29% for bilayer graphene, 35 % for 3-layer graphene and 37% for 4-layer graphene. This difference decreases at higher temperatures and can be attributed to stronger in-plane three-phonon interactions in both single and few layer graphene. Since higher phonon frequencies are involved when considering interactions of ZA modes with LA/TA modes, these are not very strong at low temperatures (due to lower occupation). Therefore, the additional interactions that appear in few-layer graphene are only important at low temperatures when the strength of ZA+ZA↔LA/TA is relatively weak. As temperature increases, the strength of 3 phonon interactions involving LA/TA modes begins to dominate those involving only ZA/ZO modes in both single- and few-layer graphene. Consequently, at high temperatures no significant difference between single and few layer graphene will exist. At 500 K, for example, the difference in thermal conductivity between single and 4-layer graphene reduces to 27%.

We note further that the computed thermal conductivity of 4-layer graphene at room temperature is 2052 W/m/K and very close to the highest reported room temperature thermal conductivity (in-plane) of pyrolytic graphite (2000 W/m.K [41].) The computations of 4-layer graphene thermal conductivity also agree very well with the measured variation of graphite thermal conductivity with temperature [41, 42]; we note that that no fitting parameters have been used in these computations. We also note that the present computational results for single-layer graphene exhibit broad quantitative agreement with recently published experimental data [21]. In Figure 3(b), the thermal conductivity values saturate at four graphene layers.

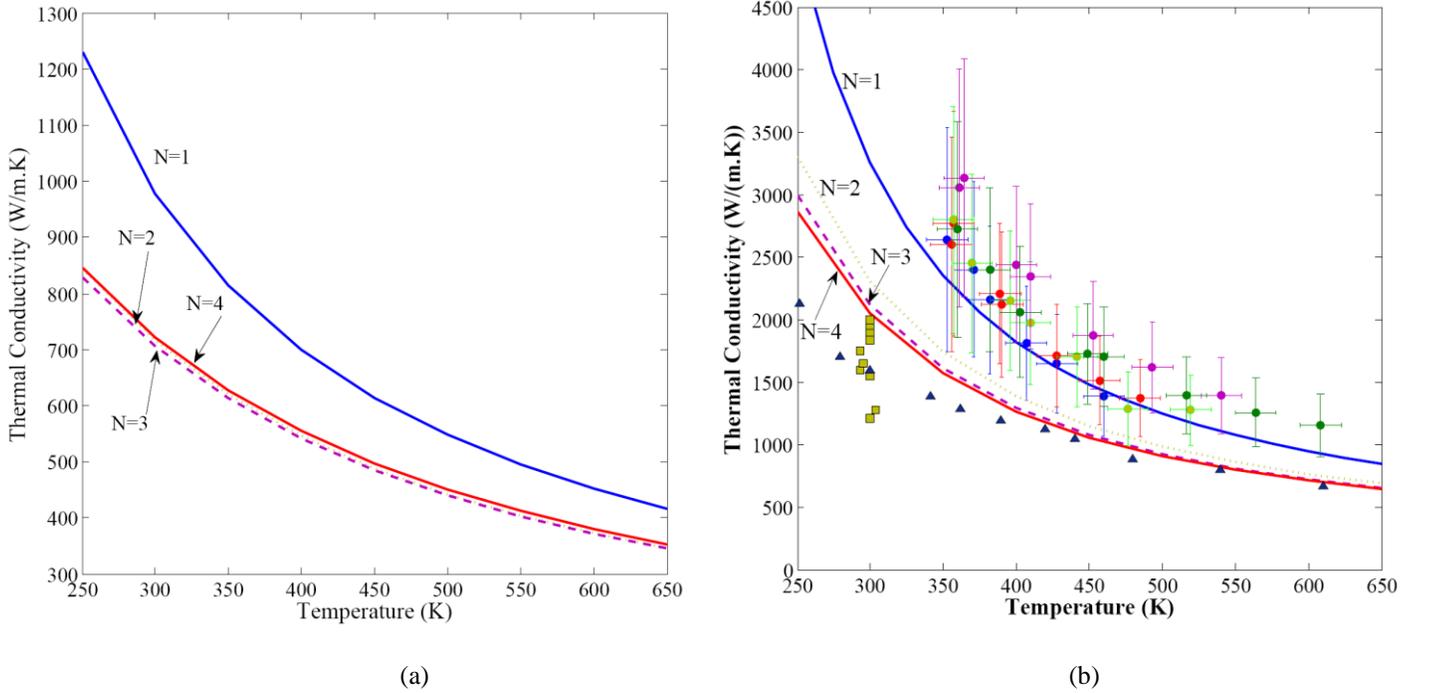

Figure 3 *Thermal conductivity of few-layer graphene along the Γ-M direction versus temperature (a) under the SMRT approximation (b) using solution of the linearized BTE The filled circles correspond to thermal measurements on single-layer graphene reported in [21] while the filled triangles correspond to thermal conductivity of graphite [42]. The filled rectangles represent the range of measured thermal conductivities in pyrolytic graphite at room temperature (from reference [41]).*

To understand the physical reason for thermal conductivity reduction with the addition of layers, we compare the case of single and bilayer graphene. Regarding single-layer graphene, it has been previously noted [12, 23, 24] that the dominant contribution to thermal conductivity is from the ZA modes and not the LA/TA modes. As explained in [12, 23], this arises due to the reflection symmetry of perfect single-layer graphene (+z is analogous to –z) which implies that third derivatives of the potential with the form $\phi_{z\beta\gamma}^{l(0)m(i)n(j)}; \alpha, \beta = x, y$ and $\phi_{zzz}^{l(0)m(i)n(j)}$ are zero. This means that only even numbers of out-of-plane phonons can participate in a three- phonon scattering. However, in bilayer (or *N* layer) graphene, this is not true as reflection symmetry is not preserved. The coupling between different layers leads to non-zero values of these third derivatives and implies that many new phonon scattering processes involving *odd* numbers of ZA/ZO phonons will become available for scattering. Additionally, we find that non-zero values of the third derivatives obtained for terms of the type $\phi_{zzz}^{l(0)m(i)n(j)}$ are much larger than derivatives of the type $\phi_{z\beta\gamma}^{l(0)m(i)n(j)}; \alpha, \beta = x, y$. These observations are true of few-layer graphene as well. Since the eigenvectors remain *almost* decoupled, these values imply that the most resistive new scattering channels in few layer graphene will involve 3 ZA/ZO phonons.

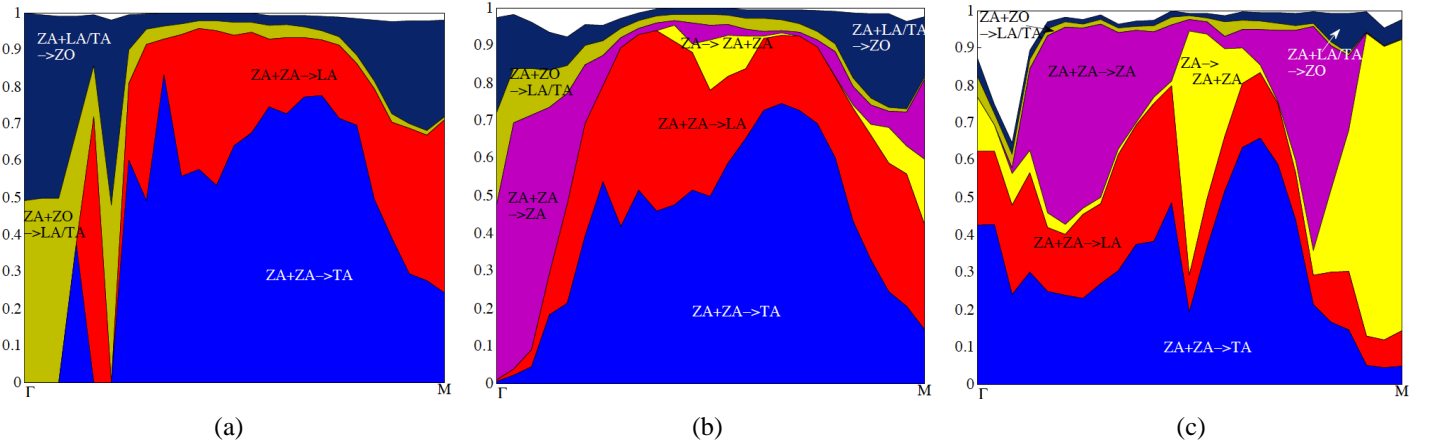

Figure 4 *Relative contribution to* $\Gamma_{\vec{k}(p)}$ *of three- phonon scattering pathways at 300 K (a) ZA mode (single layer) (b) ZA$_1$ mode (bilayer) and (c) ZA$_2$ mode (bilayer).*

Figure 4 (a)-(c) shows the relative strength of various scattering pathways for a ZA mode in single-layer graphene and the ZA and ZA$_2$ modes in bilayer graphene as a function of the wave vector magnitude along the Γ-M direction at 300 K. We use $\Gamma_{\vec{k}(p)}$ to measure the extent of phonon scattering as it gives a clear first-order picture of the relative strength of scattering events on the phonon occupation the strength of interaction (using the anharmonic IFCs) and the joint density of states for every interaction (through the d$k_l$/|$v_n$'| term). The details of the scattering processes presented in Figure 4 clearly indicate that the additional processes appearing in bilayer graphene involve three ZA phonons (from either the ZA$_1$ or ZA$_2$ branches). Furthermore, these *extra* scattering channels (compared to single-layer graphene) are important for the fundamental ZA mode only at very small wave vectors. Therefore, it is seen that the amount of scattering for the fundamental acoustic ZA$_1$ mode does not change substantially as layers are added. However, the channels $ZA_2+ZA_{1/2} \rightarrow ZA_{1/2}$ and $ZA_2 \rightarrow ZA_{1/2}+ZA_{1/2}$ are seen to contribute significant thermal resistance for the overtone (eigenvector corresponding to that shown in Figure 1(a)) throughout the BZ.

We also emphasize that many possible processes involving the degenerate phonon branches still cancel each other at finite $k$ by symmetry. Examining the matrix elements, we find that many potentially important processes such as $ZA_1+ZA_1 \leftrightarrow ZA_1$; $ZA_1+ZA_2 \leftrightarrow ZA_2$; $ZA_1+ZA_2 \leftrightarrow LA_1/TA_1$; $ZA_1+ZA_1 \leftrightarrow LA_2/TA_2$ and their permutations still contribute a zero matrix element. This implies that degeneracy does not directly increase the scattering of LA/TA modes significantly; ultimately the strength of the interactions is non-zero only for $LA_1 \leftrightarrow ZA_2+ZA_2$ and $LA_1 \leftrightarrow ZA_1+ZA_1$. Nevertheless, a few extra channels involving one ZA/ZO phonon mode appear for both LA and TA modes, which slightly increase the total scattering strength (but their contribution is negligible).

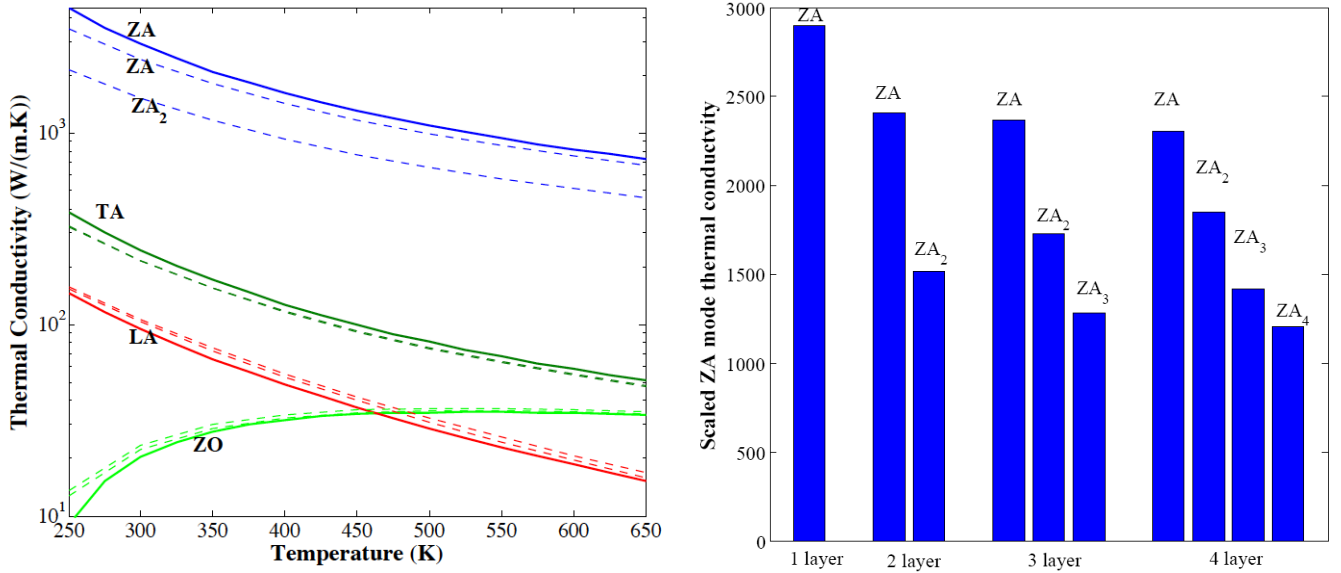

Figure 5 (a) *Scaled branch wise contribution to thermal conductivity versus temperature in single and bilayer graphene. (The solid lines correspond to single-layer graphene while the dashed lines correspond to bilayer graphene.) (b) Scaled ZA branch thermal conductivity (actual values multiplied with the number of layers) at 300 K versus number of layers.*

The contribution to thermal conductivity by the acoustic (ZA, LA, TA) and ZO modes in single and bilayer graphene are shown in Figure 5(a). The values are scaled (i.e., multiplied with the number of layers). The results indicate that any additional scattering channels significantly affect the thermal conductivity of only the higher-order overtone of the ZA mode ($ZA_2$ is the only one for bilayer graphene). Contributions by other branches (LA, TA and ZO) in bilayer graphene remain similar to those in single-layer graphene. It is also clear that both branches (i.e., the fundamental and the overtone) of the LA, TA and ZO modes contribute in the same proportion to thermal conductivity, but this not so for the ZA mode. A significant difference between the thermal conductivity of ZA and $ZA_2$ mode exists over the entire temperature range investigated here. This is a direct consequence of the weak interlayer coupling, which does not significantly alter the in-plane anharmonic IFCs to contribute more resistance.

Figure 5(b) shows the scaled contribution of the fundamental ZA branch and the higher-order overtones for 1-4 layer graphene at 300K. Most notably, the trend remains the same with the addition of the layers, i.e., the fundamental mode thermal conductivity remains comparable to that of single layer graphene (or decreases only slightly). The extent of reduction is much larger for the $ZA_2$-$ZA_4$ modes. Clearly, this result suggests that along the Γ-A direction in bulk graphite a significant decrease would exist in the ZA mode thermal conductivity (a factor of 2 suggested by our results for 4-layer graphene).

Finally, our calculations exhibit good agreement with the variation in thermal conductivity with number of layers reported in the experimental data of [16]. The measured thermal conductivities for four layers and eight layers is very similar in these experiments, suggesting that thermal conductivity saturates by four layers to the graphite value. We note that the reported thermal conductivity in [16] for 4-layer (and higher) graphene sheets is significantly lower than that of high quality bulk graphite, which implies that extrinsic factors are important in their samples – an effect not considered here. As mentioned earlier, in the theoretical calculations of [16], N processes are ignored, and any decrease in thermal conductivity is attributed to increased scattering of LA/TA modes (an artifact of the use of Klemens' matrix elements). In general, we have found that the asymptotic thermal conductivity values lie slightly below the highest reported measurements. Such a difference may easily arise from the limitations of the interatomic potentials used here. We note that the use of other force fields to describe interplanar interactions is not expected to alter the conclusions made in this paper, and a more rigorous set of anharmonic IFCs should predict a similar trend.

## VI. CONCLUSIONS

We have calculated the thermal conductivity of 1-4 layer graphene sheets by a solution of the linearized phonon Boltzmann transport equation. As for single-layer graphene, the out-of-plane acoustic modes contribute significantly to thermal conductivity and dominate conduction even with the addition of more layers. The effect of interplanar interactions is to open many new pathways for phonon scattering, most notably those involving three ZA phonons. These scattering processes significantly reduce the net thermal conductivity. The primary modification is to the overtones of the ZA modes, while the fundamental ZA mode and all other branches remain relatively unaffected. The results presented here agree very well with experimental data for both single-layer graphene and graphite, and explain the trend in experimentally observed dimensional transition of thermal conductivity with the addition of layers. The results and mechanisms illustrated here may be used in conjunction with experimental data to engineer the thermal properties of single- and few-layer graphene devices.

**ACKNOWLEDGEMENTS**

This material is based upon work partially supported by the Defense Advanced Research Projects Agency and SPAWAR Systems Center, Pacific under Contract No. N66001-09-C-2013. The authors would like to thank the research groups of Prof. Li Shi and Rodney Ruoff for generously providing the experimental data on single layer graphene. Useful discussions with Jose A. Pascual-Gutierrez are also acknowledged.

**REFERENCES**:

[1]; Novoselov, K. S.; Geim, A. K.; Morozov, S. V.; Jiang, D.; Zhang, Y.; Dubonos, S. V.; Grigorieva, I. V.; Firsov, A. A. *Science*, **2004,** 306, 666-669.

[2]Wang, X.; Ouyang, Y.; Li, X.; Wang, H.; Guo, J.; Dai, H.. *Physical Review Letters*, **2008**, 100, 206803.

[3] Balandin, A. A.; Ghosh, S; Bao, W.; Calizo, I.; Teweldebrhan, T.; Miao, F.; Lau, C. N. *Nano Letters*, **2008**, Vol. 8, 902-907.

[4] Cai, W.; Moore, A. L.; Zhu, Y.; Li, X.; Chen, S.; Shi, L.; Ruoff, R. S. *Nano Letters*, **2010**, 10, 1645-1651.

[5] Faugeras, C.; Faugeras, B.; Orlita, M.; Potemski, M.; Nair, R. R.; Geim, A. K. *ACS Nano,* **2010**, 4, 1889-1892

[6] Sutter, P. W.; Flege, J.-I.; Sutter, E. A. *Nature Materials*, 2008, 7, 406-411.

[7] Zhou, S. Y.; Gweon, G.-H.; Fedorov, A. V.; First, P. N.; de Heer, W. A.; Lee, D.-H.; Guinea, F.; Castro Neto, A. H.; Lanzara, A. *Nature Materials*, **2007**, 6, 770-775.

[8] Kedzierski, J.; Hsu, P.-L.; Healey, P.; Wyatt, P. W.; Keast, C. L.; Sprinkle, M.; Berger, C.; de Heer, W. A. *IEEE Transactions on Electron Devices*, **2008**, 55, 2078-2085.

[9] Mak, K. F.; Shan, J.; Heinz, T. F. *Physical Review Letters*, **2010,** 104, 176404.

[10] Mak, K. F.; Sfeir, M. Y.; Misewich, J. A.; Heinz, T. F. *Proceedings of the National Academy of Sciences*, **2010,** 107, 14999-15004.

[11] Zhang, Y.; Tang, T.-T.; Girit, C.; Hao, Z.; Martin, M. C.; Zettl, A.; Crommie, M. F.; Shen, Y. R.; Wang, F.; *Nature*, **2009,** 459, 820-823.

[12] Seol, J. H.; Jo, I.; Moore, A. L.; Lindsay, L.; Aitken, Z. H.; Pettes, M. T.; Li, X.; Yao, Z.; Huang, R.; Broido, D.; Mingo, N.; Ruoff, R. S.; Shi, L. *Science*, **2010**, 328, 213.

[13] Pop, E.; Mann, D.; Cao, J.; Wang, Q.; Goodson, K. E.; and Dai, H. *Physical Review Letters*, **2005**, 95

[14] Cola, B.A.; Xu, J.; Fisher, T. S. *International Journal of Heat and Mass Transfer*, 2009, **52**, 3490-3503.

[15] Huxtable, S. T.; Cahill, D. G.; Shenogin, S.; Xue, L.; Ozisik, R.; Barone, P.; Usrey, M.; Strano, M. S.; Siddons, G.; Shim, M.; Keblinski, P. *Nature Materials*, **2003,** 2, 731-734.

[16] Zhong, H. and Lukes, J. R. *Physical Review B*, **2006**, 74(12), 125403

[17] Kumar, S.; Murthy, J. Y. *Journal of Applied Physics*, **2009,** 106, 084302.

[18] Delhaes, P. (2001). *Graphite and Precursors*. CRC Press. ISBN 9056992287.

[19] Ghosh, S.; Bao, W.; Nika, D. L.; Subrina, S.; Pokatilov, E. P.; Lau, C. N.; Balandin, A. A. *Nature Materials*, **2010,** 9, 555-558.

[20] Jang, W.; Chen, Z.; Bao, W.; Lau, C. N.; Dames, C. *Nano Letters*, **2010,** 10, 3909-3913.

[21] Chen, S.; Moore, A.L.; Cai, W.; Suk, J.W.; An, J.; Mishra, C.; Amos, C.; Magnuson, C.W.; Kang, J.; Shi, L.; and Ruoff, R. S. *ACS Nano*, **2011**, 5 321–328.

[22] Wang, Z.; Xie, R.; Bui, C. T.; Liu, D.; Ni, X.; Li, B.; Thong, T. L. J. *Nano Letters*, **2011**, 11, 113-118 Article ASAP

[23] Lindsay, L.; Broido, D. A.; Mingo, Natalio *Physical Review B*, **2010**, 82, 115427.

[24] Singh, D.; Murthy, J. Y.; Fisher, T.S. *in review* **2011**

[25] Lindsay, L.; Broido, D. A.; Mingo, N. *Physical Review B*, **2010**, 161402.

[26] Pedraza, D.F. and Klemens P.G., *Carbon*, **1994**, 32, 735-741.

[27] Kong, B. D.; Paul, S.; Nardelli, M. Buongiorno; Kim, K. W. *Physical Review B*, **2009**, 80, 033406

[28] Lindsay, L.; Broido, D. A.; Mingo, N. *Physical Review B*, **2009**, 80, 125407

[29] Nika, D. L.; Pokatilov, E. P. ; Askerov, A. S. ; and Balandin A. A. *Physical Review B,* **2009**, 79, 155413.

[30] Spanu, L.; Sorella, S.; Galli, G. *Physical Review Letters*, **2009**, 103, 196401

[31] Mounet, N.; Marzari, N. *Physical Review B*, **2005**, 71, 205214.

[32] Lindsay, L.; Broido, D. A. *Physical Review B*, **2010**, 81, 205441.

[33] Girifalco, L. A.; Hodak, Miroslav; Lee, Roland S. *Physical Review B* , **2000**, 62, 13104-13110.

[34] A. A. Maradudin, E. W. Montroll, G. H. Weiss, and I. P. Ipatova, *Solid State Physics*, Suppl. 3, 2nd ed. (Academic Press, New York, 1971).

[35] Saha, S. K.; Waghmare, U. V.; Krishnamurthy, H. R.; Sood, A. K. *Physical Review B*, **2008**, 78, 165421

[36] Ziman, J. M. *Electrons and Phonons*. London, UK : Oxford University Press, 1960

[37] Pascual-Gutiérrez, José A.; Murthy, Jayathi Y.; Viskanta, Raymond. *Journal of Applied Physics*, **2009**, 106, 063532-063532.

[38] G. P. Srivastava, in *The Physics of Phonons* (Adam Hilger IOP, Bristol, 1990).

[39] Omini, M.; Sparavigna, A. *Physical Review B,* **1996**, 53, 9064-9073.

[40] A. Ward, *Ph.D. Thesis*, Boston College

[41] Uher, C., in *SpringerMaterials - The Landolt-Börnstein Database* edited by Madelung, O. and White, G. K. (http://www.springermaterials.com)

[42] Slack, Glen A. *Physical Review*, **1962**, 127, 694-701.